\documentclass[5p,sort&compress]{elsarticle}
\usepackage[utf8]{inputenc}
\usepackage{titlesec}
\usepackage{tabu}
\usepackage{xcolor}

\usepackage{lmodern}

\usepackage{hyperref}
\hypersetup{
    colorlinks=true,
    linkcolor=blue,
    filecolor=magenta,
    urlcolor=blue,
}

\usepackage{booktabs}
\usepackage{tabularx}
\usepackage{graphicx}
\usepackage{cuted}

\journal{Software Impacts}

\begin{document}

\begin{frontmatter}
\title{The Inverse Transparency Toolchain: A Fully Integrated and Quickly Deployable Data Usage Logging Infrastructure}
\author{Valentin Zieglmeier}
\affiliation{Technical University of Munich, Munich, Germany}
\ead{valentin.zieglmeier@tum.de}


\begin{abstract}
	Inverse transparency is created by making all usages of employee data visible to them.
	This requires tools that handle the logging and storage of usage information, and making logged data visible to data owners.
	For research and teaching contexts that integrate inverse transparency, creating this required infrastructure can be challenging.
	The Inverse Transparency Toolchain presents a flexible solution for such scenarios.
	It can be easily deployed and is tightly integrated.
	With it, we successfully handled use cases covering empirical studies with users, prototyping in university courses, and experimentation with our industry partner.
\end{abstract}

\end{frontmatter}

\newcommand{\githubRepoUrl}{https://github.com/SoftwareImpacts/SIMPAC-2023-126}

\begin{strip}

	\vspace{-2.75em}

	\footnotesize
	\begin{tabularx}{\textwidth}{ X p{9.5cm} }
		\toprule
		Current code version & v1.0.0 \\
		Permanent link to code/repository used for this code version & \url{\githubRepoUrl} \\
		Legal Code License   & MIT License \\
		Code versioning system used & Git \\
		Software code languages, tools, and services used & Python, Go, Svelte, JavaScript, Shell; HashiCorp Vault, HashiCorp Consul, SQLite, Docker\\
		Operating environments \& dependencies & Python $\geq$ 3.8, Go $\geq$ 1.14, Ubuntu $\geq$ 20.04; dependencies: \url{\githubRepoUrl/blob/master/DEPENDENCIES.md}\\
		Link to developer documentation/manual & \url{\githubRepoUrl/blob/master/README.md} \\
		Support email for questions & \href{mailto:valentin.zieglmeier@tum.de}{valentin.zieglmeier@tum.de}\\
		\bottomrule
	\end{tabularx}

	\vspace{3em}

\end{strip}

\section{Introduction}
Employees become increasingly transparent in the digital workplace.
Every aspect of work, including the collaboration with colleagues, management of vacation or sick days, but importantly also the status and results of the work itself can be tracked in the tools that are used for these tasks.
This transparency can be necessary to run the business or it may simply facilitate work and collaboration, but it is unbalanced.
Managers have a much clearer picture of the conducted work, but employees have no tools to oversee how their data are used or uncover potential misusage of their data.
Therefore, the idea of creating \emph{inverse transparency}~\cite{brin1998transparent} in the workplace was proposed~\cite{gierlich2020more, zieglmeier2021trustworthy}.
In short, this entails making all usages of employees' data transparent to them.
Thereby, misusage of data could be addressed by enabling accountability.~\cite{zieglmeier2023itbd}

To realize inverse transparency, a backbone of auxiliary tools is required.
These are necessary to handle the technical steps that enable inverse transparency: logging and storing usage information from various tools and making those logged data visible to data owners.
To create a fully integrated setup, a single-sign on system for user authentication and attribution is additionally needed.~\cite{zieglmeier2021trustworthy}
Especially for research projects and university courses that need to quickly set up this required infrastructure, the overhead that this setup entails can endanger the success of projects.
But, also when introducing inverse transparency in practice, having a basic setup for experimentation that allows easy extension and integration is vital.
This enables small-scale trials and studies without needing to invest in a fully fledged system up front.

\begin{figure*}[t]
	\centering
	\includegraphics[width=.85\linewidth]{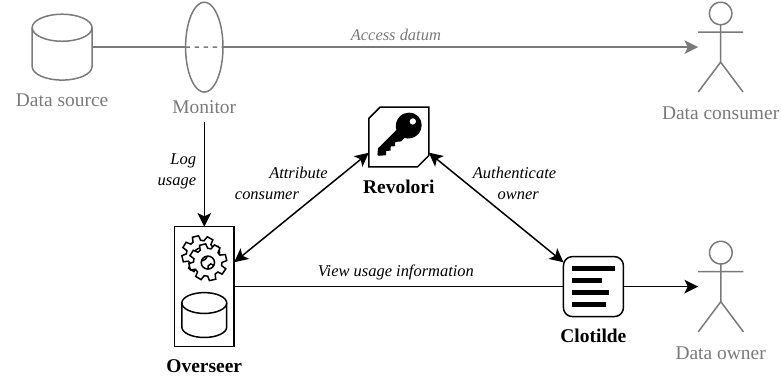}
	\caption{Overview of how the toolchain components interact. Arrows denote interactions and data flow. Grayed out components are not part of the toolchain, instead representing its operating context.}
	\label{fig:toolchain-overview}
\end{figure*}

With the Inverse Transparency Toolchain, we present a solution for all these cases.
As a fully integrated system, it can be easily deployed in minutes, requiring only basic configuration.
This enables research projects, theses, or university courses to be set up quickly.
Furthermore, its modular architecture of individual tools means that any component can be replaced or adapted individually, enabling, e.g., experimentation with different variations or a step-by-step rollout of larger setups.
In our own work, we have already successfully applied the toolchain in such scenarios.
In the context of research projects and user studies, it allowed us to depend on a robust backbone that ensured a fully inverse transparent environment.
For students working on theses or in courses, the toolchain offered a standardized interface that allowed them to integrate their projects with ease, sparing them the effort of creating and setting up their own tools from the ground up.
This meant that they could get started on their projects almost immediately.
Finally, in our work with our industry partner, employees worked with the toolchain to prototype and experiment with their own solutions.
Our choice of creating a toolchain of web apps meant that they did not need to set up a local infrastructure and could focus on developing valuable use cases for their context.

\section{Tools and Functionalities}

Technically realizing inverse transparency, at its core, requires three steps: monitoring and attributing data usages, storing the created logs, and making those data available to view~\cite{zieglmeier2021trustworthy}.
When considering these tasks, we find that inverse transparency solutions for different contexts vary fundamentally only in one aspect, namely in \emph{how} and \emph{what} is monitored.
That is a domain- and context-specific task that cannot easily be abstracted away.
To integrate monitoring into data usage processes, we envision building the analytics with \emph{inverse transparency by design}~\cite{zieglmeier2023itbd}.
To support this vision, our toolchain provides tools for exactly those tasks that can be abstracted away: \emph{Overseer} logs usages and stores them, \emph{Clotilde} displays stored usage data, and \emph{Revolori} authenticates and attributes users (see \autoref{fig:toolchain-overview}).
In the following, we give a brief overview of each tool individually.

\subsection{Overseer: logs and stores data usages}
As the interface of inverse transparency monitors to the data store, \emph{Overseer} can be thought of as the backend of the toolchain.
It enables read and write access to the usage log data via a REST API.
It is implemented in Python utilizing FastAPI.\footnote{\url{https://fastapi.tiangolo.com}}
That means that an OpenAPI\footnote{\url{https://www.openapis.org}} documentation is automatically created and hosted when running it.
Data are stored in a local SQLite\footnote{\url{https://sqlite.org}} database file, making the system very portable and easy to set up, using the the SQLalchemy\footnote{\url{https://www.sqlalchemy.org}} ORM.
If the database schemas or models are modified, \emph{Overseer} automatically migrates its database with Alembic.\footnote{\url{https://alembic.sqlalchemy.org}}
Finally, should that be necessary, the implemented \emph{data access object} pattern facilitates replacing the storage backend.

\subsection{Clotilde: makes logged usage information accessible}
Arguably the most important component is \emph{Clotilde}, which can be thought of as the frontend of the toolchain.
It is a modern single-page web application implemented in JavaScript with Svelte,\footnote{\url{https://svelte.dev}} using the Rollup\footnote{\url{https://rollupjs.org}} module bundler for deployment.
The user interface styling is based on Bootstrap.\footnote{\url{https://getbootstrap.com}}
Through \emph{Clotilde}, users can access any usage logs relating to their data.
A summary is provided to give an overview over data usages in the past seven days.
Furthermore, a full table of all logged data usages is available.
The viewed data can be exported in form of a PDF report.
To allow users to limit who uses their data, a page for simple usage policies exists.
It is relatively elementary, but can be easily extended to explore various options for a usage policy screen.
If this functionality is not required, the page can easily be hidden from the navigation bar.

\subsection{Revolori: authenticates and attributes users}
Both \emph{Overseer} and \emph{Clotilde} depend on \emph{Revolori}, a single sign-on server that offers both user authentication but also attribution services.
It is a lightweight implementation in Go, depending on HashiCorp Vault\footnote{\url{https://www.vaultproject.io}} for secret management and encryption, which in turn utilizes the HashiCorp Consul Secrets Engine.\footnote{\url{https://developer.hashicorp.com/vault/docs/secrets/consul}}
The secrets engine can easily be replaced, but even Vault can be supplanted by a different system, if necessary.
\emph{Revolori} offers a REST API.
We integrate the Swaggo swag library\footnote{\url{https://github.com/swaggo/swag}} to automatically create and host an OpenAPI documentation when running it.
As its main task, the \emph{Revolori} API offers endpoints for user session management, enabling login and logout functionality.
To achieve that, JSON web tokens\footnote{\url{https://jwt.io}} are issued and tracked in the system.
This functionality is used by \emph{Clotilde} to log in users, but can also be integrated into inverse transparency monitors.
Its secondary task, used by \emph{Overseer}, is user attribution.
To motivate this briefly, individuals use various identifiers depending on the tool they interact with: often their primary email, but they may have multiple emails or services may expect a username.
Single sign-on integration can help in most cases, but some tools, such as Git, do not support it.
To identify users and attribute data usages correctly, Revolori can store all related identifiers of users and provides an API that matches any given identifier to their main identifier.

\section{Usage Examples}

The flexibility of the toolchain allows for various usage scenarios.
In the following, we give a brief overview of some example applications.

\subsection{Base case: Using the integrated toolchain}

As the primary use case, the toolchain can be used unchanged, with various usage monitors feeding data into Overseer (see \autoref{fig:toolchain-overview}), enabling \emph{inverse transparency}.
For example, in a study, we let developers implement people analytics with inverse transparency by design~\cite[see][]{zieglmeier2023itbd}.
To realize inverse transparency, the developed people analytics only needed to call the Overseer API to log usage information.
Overseer stored the usage logs in a central database for later access, without the developers needing to implement this functionality for each tool.
Then, data owners could use Clotilde to view the logged information, gaining inverse transparency over how their data were used~\cite[see][Sec.~5.2]{zieglmeier2023itbd}.

\subsection{Integrating a custom dashboard with Overseer and\\Revolori}

One of the important advantages of the toolchain is its flexibility, as it offers the ability to replace individual components.
A recent research field that can benefit from this adaptation is the area of \emph{privacy dashboards}~\cite[e.g.,][]{angulo2015usable, bier2016privacyinsight, herder2020privacy, farke2021privacy}.
These are dashboards that display personal data flows, among other information.
Clotilde can be considered a basic instantiation of a privacy dashboard.
Researchers working on their own privacy dashboard can simply replace Clotilde with their own dashboard, and benefit from the robust infrastructure of the toolchain for data storage and user authentication.
This removes the need to implement their custom solution~\cite[see, e.g.,][Sec.~5]{bier2016privacyinsight} and accelerates the research.

\subsection{Adapting Clotilde to test user interface modifications}

Some research does not require a full privacy dashboard, but is limited to individual features and qualities.
For example, researchers consider usability, user experience, or trustworthiness of transparency tools and dashboards~\cite[see][]{earp2016had, raschke2018designing, zieglmeier2021designing}.
To that end, the extensibility of Clotilde can be utilized to test user interface modifications.
We specifically designed the tool to be modular, meaning individual pages and components can be replaced.
We made use of this in our own research, designing two variants of the dashboard: one to elicit trust, one to reduce it.
This allowed us to quickly evaluate the effectiveness of our measures in influencing user trust~\cite[see][Sec.~6]{zieglmeier2021designing}.

\section{Impact Overview}

The relatively large overhead of providing the auxiliary tools required for an inverse transparent environment limits the resources that remain to create truly novel solutions for inverse transparency.
This is the core issue that is addressed by the Inverse Transparency Toolchain.
Thereby, it supports researchers, lecturers, and practitioners, enabling use cases that would otherwise be infeasible to address.

For researchers, the modular design of the toolchain allows focusing on only those details that are relevant for their research question.
As one example, many open research questions still exist with regard to how an effective inverse transparency dashboard should be designed.
One needs to balance the ideal of transparency, which would suggest making available all information, and the understandability and usability of the information, which may mean simplifying, aggregating, or interpreting for users.
Being able to use the existing toolchain and only modifying the user interface means that realistic exploration and user studies are feasible with minimal adaptations to the code.
In our own research, we employed the Inverse Transparency Toolchain in varying ways, which lead to multiple published and in-review papers: \cite{zieglmeier2021trustworthy, zieglmeier2021designing, gierlich2021leading, zieglmeier2023itbd, zieglmeier2023decentralized}

For lecturers, the simple and quick deployment of the toolchain enables them to teach courses that integrate inverse transparency.
On the one hand, this can mean letting students experience inverse transparency and empowering them to consider what types of data usages they find acceptable.
On the other hand, especially in computer science, students can be guided to build their own inverse transparency solutions, e.g. focusing on developing domain-specific monitors or creating analysis tools with inverse transparency by design.
We conducted two university practicums exactly for this purpose.
Students built various analysis tools that integrated inverse transparency, showing the different ways that it can impact the software design and architecture.

Finally, practitioners can benefit when bootstrapping inverse transparency pilots or for experimentation.
Due to inverse transparency focusing on improving employees' knowledge about how their data are used, such projects may be driven by grassroots efforts from within the company.
The out-of-the-box deployment and tight integration of the Inverse Transparency Toolchain allows even single individuals or small teams to set up a proof of concept infrastructure.
If resources are available, the web-based approach eases experimentation, as individuals are not required to install tools on their machine to test out inverse transparency tools they develop.
In our work with our industry partner, the toolchain enabled employees' experimentation and development of own use cases~\cite[see][pp.~70--73]{boes2022daten}.
This work was conducted as part of the research project ``Inverse Transparenz,''\footnote{\url{https://www.inversetransparenz.de/}} which ran from 2018 to 2022.
The Inverse Transparency Toolchain is one of our results.

\section{Limitations and Potential Improvements}

The toolchain was developed primarily for research and teaching purposes. This focus lead to some limitations, which we identify in the following.

While all \emph{Overseer} API endpoints for reading usage data or manipulating access policies use \emph{Revolori} authentication, we chose to only require basic authentication for the API endpoints used to add new usage logs.
This was done to simplify development of new inverse transparency monitors, as implementing a complex authentication flow does not improve the core functionality of such tools but costs time and effort.
Yet, this means that \emph{Overseer} trusts the monitors to ensure the correctness of log entries.
Furthermore, the data stored in its SQLite database is not encrypted, which allows server admins to read, modify, and delete all entries.
This was done to enable the manipulation of log entries for specific use cases, e.g. in studies, but means that the integrity of log entries cannot be guaranteed.
Therefore, one potential step for improvement would be to focus on providing integrity guarantees for the data stored by \emph{Overseer}.

Furthermore, we made sure that \emph{Revolori} stores user data securely, utilizing the established HashiCorp Vault system.
Yet, our focus was on security, meaning the usability is not optimized for.
Specifically, user registration has to be done manually via the API, which may be a hurdle for non-technical users.
The OpenAPI website \emph{Revolori} provides does make this simpler and provides some form of user interface, but it is not ideal.
Furthermore, to simplify user management, the admin user can currently access and manipulate all user data, except for passwords.
To improve this, users could be provided a self-service interface to create accounts, manage their data, or delete it.
This would make the admin privileges obsolete and thereby improve information security.

\section{Conclusion}

Inverse transparency presents a new way to think about data privacy in the workplace, proposing to allow employees to monitor how their data are used.
This presents interesting use cases for researchers, lecturers, and practitioners.
Yet, the large setup overhead required before new challenges can be addressed has the potential to hinder innovative ideas.
Therefore, we provide the Inverse Transparency Toolchain, a flexible solution for setting up an inverse transparent environment in minutes.
It offers tools for the basic tasks of inverse transparency, with a modular setup that allows replacing or complementing the provided tools depending on the use case.
With it, we hope to support novel research on inverse transparency, innovative courses, but also initiatives that aim to bring inverse transparency into practice.

\sloppy
\section*{Acknowledgments}

Supporting developers (listed alphabetically): Yiyu Gu, Johannes Heilmann, Raphael Hohmann, Stefan Knilling, Stefan Madzharov, Patipon Riebpradit, Felix Schorer. Translations contributed by Gabriel Loyola Daiqui.

This work was supported by the German Federal Ministry of Education and Research (BMBF) under grant no. 5091121.

\bibliographystyle{elsarticle-num}
\bibliography{references}
\fussy

\end{document}